\documentstyle[12pt]{article}
 1
\def\citenum#1{{\def\@cite##1##2{##1}\cite{#1}}}
\def\citea#1{\@cite{#1}{}}

\def\Om{\Omega(s,b)}

\def\dm{\frac{d\sigma}{dM^{2}dt}}
\def\dm2{\frac{d\sigma}{dM^{2}}}

\def\beq{\begin{equation}}
\def\eeq{\end{equation}}
\def\bea{\begin{eqnarray}}
\def\eea{\end{eqnarray}}
\def\underarrow#1{\mathrel{\mathop{\longrightarrow}\limits_{#1}}}
\def\bbbz{{\mathchoice {\hbox{$\sf\textstyle Z\kern-0.4em Z$}}
{\hbox{$\sf\textstyle Z\kern-0.4em Z$}}
{\hbox{$\sf\scriptstyle Z\kern-0.3em Z$}}
{\hbox{$\sf\scriptscriptstyle Z\kern-0.2em Z$}}}}
\relax
\begin{document}
\begin{titlepage}
\noindent
\begin{flushright}
 February  1995    \\TAUP 2235-95  \\
CBPF FN - 009/95                    \\[9ex]
\end{flushright}
\begin{center}
{\Large \bf Effects of Shadowing in       }   \\[1.4ex]
{\Large \bf Double Pomeron Exchange Processes } \\ [11ex]

{\large E.\ Gotsman$^{1)}$                        }    \\[1ex]
{\large E.M.\ Levin  $^{ 2),3),a)} $                          }    \\[1ex]
{\large U.\ Maor  $^{  1),3)}$               }    \\[1.5ex]
{$^{1)}$  School of Physics and Astronomy  }      \\
{Raymond and Beverly Sackler Faculty of Exact Sciences} \\
{Tel Aviv University, Tel Aviv 69978, Israel}  \\ [1.5ex]
 {$^{2)}$   Mortimer and Raymond Sackler Institute of Advanced Studies}\\
{School of Physics and Astronomy, Tel Aviv University}\\
{Tel  Aviv, 69978, Israel}\\ [1.5ex]

{ $^{3)}$ LAFEX, Centro Brasileiro de Pesquisas F\'\i sicas / CNPq}\\
{Rua Dr. Xavier Sigaud 150, 22290 - 180, Rio de Janiero, RJ, Brasil}

\footnotetext{ a) On leave from Petersburg Nuclear Institute, 188350, Gatchina,
St. Petersburg, Russia}
\end{center}
{\large \bf Abstract:} The effects of shadowing  in double Pomeron exchange
processes are investigated within an eikonal approach with a Gaussian input.
Damping factors due to screening are calculated for this process and
compared with the factors obtained
 for total, elastic and single diffraction cross sections.
Our main conclusion is that  counting rate calculations,
of various double Pomeron exchange processes (without screening
 corrections) such as heavy quark and Higgs production
 are reduced by a factor of 5 in the LHC energy range, when
screening corrections are applied.
\begin{quotation}
\end{quotation}
\end{titlepage}
%%%%% END TITLE PAGE
\par
   The process of double Pomeron exchange (DPE),  shown in Fig.1, has
been recognized for sometime [1,2] as an interesting window through
which we can further persue our study of Pomeron dynamics, and
extend our knowledge of diffraction. Even though DPE processes have
relatively small cross sections, they have a very clean experimental
signature, where the central diffractive cluster is seperated
from the remnants of the two projectiles by large rapidity gaps.
(For a schematic lego plot see Fig.2). DPE processes have
recently attracted a considerable amount of attention as a
possible background for rare electroweak events \cite{3},
as well as actual sources for central diffraction of
$ q \bar{q}$ jets \cite{4} and minijets \cite{5},
   heavy flavor production \cite{6},
Higgs production \cite{7} and an interesting configuration
for the study of the Pomeron structure \cite{8}.
\par
    In this paper we wish to examine the consequences of including
screening corrections in the initial state of DPE diagrams.
Our calculations are applicable to DPE calculated
either in the conventional Regge formalism, or through a two gluon
exchange approximation \cite{9,10}.
We  show that these s-channel unitarity corrections, cause DPE
processes to be strongly suppressed throughout the Tevatron energy
range, and even more so at higher energies. The degree of this
suppression can easily be assessed in terms of a damping factor
$ < \vert D \vert ^{2} > $. In this paper we proceed
to calculate the damping factor for DPE processes, and make
a realistic evaluation of some of the associated final states.
As we shall show, our estimates are   considerably
smaller than the  uncorrected  rates, published previously.
\par
     The present investigation extends our ongoing study \cite{11}
on the implementation of s-channel unitarity corrections to Pomeron
exchange in high energy hadron scattering. Our study has mostly
concentrated on a supercritical DL soft Pomeron \cite{12}
$$ \alpha(t) = 1 + \Delta + \alpha^{\prime} t $$
with $ \Delta \simeq 0.085 $ and $ \alpha^{\prime}\simeq  0.25\; GeV^{-2} $.
This simple model is rather successful in reproducing the available
hadronic data on total and elastic cross sections.
We note, nevetheless, that our method is equally effective if we
choose to calculate with the hard BFKL QCD-Pomeron \cite{13}.
This will be the main subject of a paper to be published shortly.
\par
    In previous publications \cite{11} we have attempted
a systematic study of s-channel unitarity screening corrections
in an eikonal approximation, where our b-space amplitude
is written as
 \begin{equation}
a(s,b) = i ( 1 - e^{- \Omega(s,b)} )
\end{equation}
To obtain analytic expressions for the cross sections
of interest, we make the following simplifying assumptions:

1) The opacity $ \Omega(s,b) $ is a real function,  i.e. a(s,b) is
pure imaginary, when necessary analyticity and crossing symmetry
can be easily restored \cite{11}.

2) We assume our input opacity to be a Gaussian
 \begin{equation}
 \Om =\nu(s) e^{-\frac{b^{2}}{R^{2}(s)}}
\end{equation}
which corresponds to an exponential behaviour of the input
amplitude in t-space. For a supercritical amplitude we have
 \beq
\nu(s) = \frac{\sigma_{0}}{2 \pi R^{2}(s)}(\frac{s}{s_{0}})^{\Delta}
  \eeq
where
\beq
 R^{2}(s) = 4 [ R^{2}_{0} + \alpha^{\prime} ln \frac{s}{s_{0}} ]
\eeq
and  $ \sigma_{0} = \sigma(s_{0}) $ .

3) Our eikonal approximation does not include diffractive rescattering.
Neglecting these states is a reasonable approximation,
as $ \frac{\sigma_{diff}}{\sigma_{inel}} \ll $ 1 throughout the
energy domain of interest.

\par
   The simplified model,  having the properties described above,
reproduces the main features
of the elastic and diffractive channels under
investigation. To obtain an estimate of the suppression
induced by s-channel unitarity screening corrections, we
define a damping factor $ < \vert D_{i} \vert^{2} > $,
which is the ratio of the eikonalized output cross section
of interest to the uncorrected input cross section.
{}From the definitions of $ \sigma_{tot} $ and $ \sigma_{el}$
\cite{11} we have
\beq
< \vert D_{tot} \vert ^{2}> =
\frac{ \sigma^{out}_{tot}}{\sigma^{in}_{tot}}
= 1 - \sum_{n=1}^{\infty} (-1)^{n + 1}
\frac{ \nu^{n}}{(n+1)^{2} n!}
\eeq
\beq
< \vert D_{el} \vert ^{2}> =
\frac{ \sigma^{out}_{el}}{\sigma^{in}_{el}}
= 1 -   4\, \sum_{n=1}^{\infty} (-1)^{n + 1}
\frac{ \nu^{n} [\, 2 ^{n + 1} \,-\,1\,]}{(n+2)^{2} ( n + 1 )!}
\eeq
For the inelastic channels, the damping factor is defined
\beq
< \vert D_{i} \vert^{2} >
= \frac{\int d^{2}b\; a_{i}(s,b) \; P(s,b)}
{\int d^{2}b \;a_{i}(s,b)}
\eeq
where $ a_{i}(s,b) $  is the b-space amplitude of interest, and
$ P(s,b) = e^{-2 \Omega(s,b)} $ denotes the probability \cite{11}
that no inelastic interaction takes place at impact parameter
b. We note that the definition of
$ < \vert D_{i}^{2} \vert > $ is correlated to the definition
of the survival probability $ < \vert S \vert^{2} > $
in the case of hard parton scattering \cite{3}.
\par
    We would like to mention that the physical meaning
of the damping factor in the parton approach, is the probability
to have one parton shower collision in the hadron-hadron
interaction.  Bjorken's survival probability is the damping
factor multiplied by the ratio of the input cross section
to the inclusive one, with the same trigger in the one parton
shower collision. For example the Bjorken survival probability
for high $ p_{T} $ jet central diffraction is
\beq
< \vert S \vert^{2} > = < \vert D \vert^{2} > \cdot
\frac{\sigma^{one\; parton\; shower}(high \;p_{T}\; jet\; in\; double
\;pomeron \; collision)}{\sigma_{inclusive}(high\; p_{T}\; jet)}
\eeq
For many reactions the second factor has been calculated, for some
even the denominator has  been measured. This is the reason
for our interest in calculating the damping factor.
\par
 As we have seen \cite{11}, the damping factor for single
diffractive dissociation (SD), (calculated in the triple
Pomeron limit), is
\beq
< \vert D_{SD} \vert^{2} > =
\frac{(M^{2}\dm2)^{out}}{(M^{2}\dm2)^{in}}
= a (\frac{1}{2\nu})^{a} \gamma(a,2 \nu)
= 1 - a\, \sum _{n=1}^{\infty}(-1)^{n + 1} \frac{(2\nu)^{n}}{(a+n)n!}
\eeq
where
 \beq
 \frac{4}{3} \,\,<\,\,
a(s,M^{2}) = \frac{2 R^{2}(s)}{ {\bar R}^{2}(\frac{s}{M^{2}}) +
     2 {\bar R}^{2}(\frac{M^{2}}{s_{0}})}  \leq 2
\eeq
and
 \beq
    {\bar R^{2}}(\frac{s}{M^{2}}) = 2 R^{2}_{0} + r^{2}_{0}
      + 4 \alpha^{ \prime} ln(\frac{s}{M^{2}})
 \eeq
   $ r_{0} \ll R^{2}_{0} $ denotes the radius of the triple
 vertex and can be neglected. \\
$ \gamma(a,2 \nu) = \int^{2 \nu}_{0} z^{a - 1} e^{-z} dz $ ,
  denotes the incomplete Euler gamma function. One has to
integrate over $ M^{2} $ to obtain the
 integrated SD cross section $ \sigma_{SD} $.
We find that $a(s,M^{2})$ has a rather weak dependence on $M^{2}$
as it is proportional to
  $ \alpha' ln M^2$
 ( with  $\alpha'$ small) over a relatively narrow domain.
 So in practice, one can factor out
$ < \vert D_{SD} \vert^{2} > $  in the
$ M^{2} $ integration.
 Thus we have
 \beq
\frac{\sigma^{out}_{SD}}{\sigma^{in}_{SD}}
     \simeq  < \vert D_{SD} \vert ^{2} >
\eeq
\par
    In the exceedingly high energy limit $ \nu(s) \gg $ 1 and
 $ a(s,M^{2}) \rightarrow $ 2 from below. In this limit the
 screened cross sections differ drastically from the pole  input
as is evident from Table I. However, in the energy range that is
of interest,
 i.e. HERA-Tevatron-LHC, $ \nu(s) $ appears to be of the order of unity
(from estimates of
screening effect from the available experimental data ).
For a more realistic estimate we list in Table II the different damping
factors at some typical accelerator energies. The calculation is based
on a DL input \cite{12} of $\Delta$ = 0.085,  $\alpha'$ = 0.25 $GeV^{-2}$
and $R^2_0$ = 5.2 $GeV^{-2}$.

   We note that the screening corrections saturate at different
energy scales for the different channels. In particular, diffractive
channels, such as $ p\;p \rightarrow p\;X $ and even more so
$ \gamma \; p \rightarrow \psi \;X $, for which $ a(s) \rightarrow $
 2 precociously, exhibit a very tempered energy dependence, which
is the result of the early saturation of the screening corrections.
 This behaviour is to be contrasted with the effective power behaviour
of the total and elastic cross sections at these energies.
\par
   We now turn to a detailed calculation of central diffraction (CD),
which proceeds through DPE. This process was originally calculated
by Streng \cite{2}. We follow this calculation and then proceed  to
calculate $ < \vert D_{CD} \vert^{2} > $. The relevant kinematics
are shown in Fig.3. We remind the reader that the t-space elastic
scattering amplitude is given by
\beq
    F(s,t) = \frac{1}{2} \nu(s) R^{2}(s) e^{B(s)t}
\eeq
  where the slope of the amplitude $ B(s) =\frac{R^{2}}{4} $.
   Adopting Streng's notation \cite{2}  we have
\beq
    F(s,t) = 2 g^{2}_{P}(0) (\frac{s}{s_{0}})^{\Delta}
                e^{B(s)t}
\eeq
\beq
    \nu(s) = \frac{4 g^{2}_{P}(0)}{R^{2}(s)} (\frac{s}{s_{0}})^{\Delta}
\eeq
    In this notation
\beq
  \sigma_{tot} =4 \pi F(s,0) = 8 \pi g^{2}_{P}(0)
\eeq
The cross section of interest is given by
\begin{eqnarray}
 s_{1}s_{2} \frac{d^{4}\sigma}{ds_{1}ds_{2}dt_{1}dt_{2}}
  &=&    4 \pi^{2} \frac{\sigma_{PP}(M^{2})}{\sigma^{2}_{0}} \cdot
       F[\frac{s}{s_{1}},(\frac{q}{2} + k_{1})^{2}] \cdot
   F[\frac{s}{s_{1}},(\frac{q}{2} - k_{1})^{2}] \cdot \nonumber \\
& &   F[\frac{s}{s_{2}},(\frac{q}{2} + k_{2})^{2}] \cdot
 F[\frac{s}{s_{2}},(\frac{q}{2} - k_{2})^{2}]
\end{eqnarray}
 where we have used Streng's definition of $ \sigma_{PP} $.
 Eq. (17) can be rewritten as ($t = - q^2$)
\begin{eqnarray}
 s_{1}s_{2} \frac{d^{4}\sigma}{ds_{1}ds_{2}dt_{1}dt_{2}}
    & =&  4 \pi^{2} \frac{\sigma_{PP}(M^{2})}{\sigma^{2}_{0}} \cdot
 (\frac{1}{4})^{4} {\bar R}^{4}(\frac{s}{s_{1}})
{\bar R}^{4}(\frac{s}{s_{2}}) {\bar \nu}^{2}(\frac{s}{s_{1}})
 {\bar \nu}^{2}(\frac{s}{s_{2}}) \cdot  \nonumber \\
    &  &    e^{ -\,{\bar B}(\frac{s}{s_{1}})(2 k^{2}_{1} + \frac{q^{2}}{2})}
e^{-\,{\bar B}(\frac{s}{s_{2}})(2 k^{2}_{2} + \frac{q^{2}}{2})}
\end{eqnarray}
 where
 \beq
    {\bar B}(\frac{s}{s_{i}}) = \frac{1}{2} R^{2}_{0}
      +  \alpha^{ \prime} ln(\frac{s}{s_{i}})
= \frac{{\bar R}^{2}_{i} (\frac{s}{s_{i}})}{4}
 \eeq
\beq
{\bar \nu}(\frac{s}{s_{i}}) = \frac{\sigma_{0}}{2 \pi {\bar R}
  (\frac{s}{s_{i}})} (\frac{s}{s_{i}})^{\Delta}
 \eeq
Note that in Eq.(18) we have a factor 4 in the denominator. This is
in accord with the Reggeon calculus rules for identical particles.
 After integrating over $ k^{2}_{1}$ and $ k^{2}_{2} $ we have
\beq
 s_{1}s_{2} \frac{d^{2}\sigma}{ds_{1}ds_{2}} =
 \frac{4 \sigma_{PP}(M^{2})}
 {{\bar R}^{2}(\frac{s}{s_{1}})
 {\bar R}^{2}(\frac{s}{s_{2}})} \cdot
 g^{4}_{P}(0) (\frac{s}{M^{2}})^{2 \Delta} \cdot
 e^{-\,\frac{1}{2}(R^{2}_{0} + \alpha^{\prime} ln\frac{s}{M^{2}})q^{2}}
\eeq
  Since $ M^{2} < s_{1} < s $ and $ s_{2} =  \frac{M^{2}s}{s_{1}} $
we can integrate over $ s_{1} $  and obtain
\begin{eqnarray}
   M^{2} \frac{d \sigma}{d M^{2}}
  &=&      \frac{1}{2 \alpha^{\prime}} \sigma_{PP}(M^{2})g^{4}_{0}
         (\frac{s}{M^{2}})^{2 \Delta}
 e^{-\,\frac{1}{2}(R^{2}_{0} + \alpha^{\prime} ln\frac{s}{M^{2}})q^{2}}
 \frac{1}{R^{2}_{0} + \alpha^{\prime}ln\frac{s}{M^{2}}} \nonumber \\
& &\cdot ln [ \frac{R^{2}_{0} + 2 \alpha^{\prime} ln\frac{s}{M^{2}}}
    {R^{2}_{0}}]
\end{eqnarray}
We define
\beq
F_{PP}(s,q^{2}) = \dm2 \frac{1}{4 \pi}
\eeq
\beq
 \sigma^{PP}_{tot}(s) = 4 \pi\; Im F_{PP}(s,q^{2} =0)
\eeq
which is identical to the cross section orginally calculated
  by Streng \cite{2}.
\par
    The above cross section grows like $ s^{2 \Delta} $ with energy,
much faster than $ \sigma_{tot} $, so that s-channel unitarity
 corrections are necessary. We proceed to calculate these
in a manner analogous to our previously published SD calculations
\cite {11}. For this purpose we write
\begin{eqnarray}
 \Omega_{PP}(s,b)
   &=&   \frac{1}{8 \pi \alpha^{\prime}} \sigma_{PP}(M^{2})g^{4}_{0}
         (\frac{s}{M^{2}})^{2 \Delta}
           \frac{1}{R^{2}_{0} + \alpha^{\prime}ln\frac{s}{M^{2}}}
 ln [ \frac{R^{2}_{0} + 2 \alpha^{\prime} ln\frac{s}{M^{2}}}{R^{2}_{0}}]
 \nonumber        \\
& & \cdot \frac{1}{2 \pi} \int d^{2}q e^{-i(q.b)}
 e^{-\,\frac{1}{2}(R^{2}_{0} + \alpha^{\prime} ln\frac{s}{M^{2}})q^{2}}
\end{eqnarray}
which can be rewritten as
\begin{eqnarray}
\Omega_{PP}(s,b) = \frac{1}{8 \pi \alpha^{\prime}}
  ln [\frac{{\bar R}^{2}(\frac{s}{M^{2}})}{{\bar R}^{2}(1)}] \cdot
   \nu^{2}(\frac{s}{M^{2}})\,\, e^{- \frac{2 b^{2}}{R^{2}(\frac{s}{M^{2}})}}
\end{eqnarray}
The screened expression for $ M^{2}\dm2 $ is then
\begin{eqnarray}
M^{2} \dm2 =
 \frac{1}{8 \pi \alpha^{\prime}} \sigma_{PP}(M^{2})
  ln [\frac{{\bar R}^{2}(\frac{s}{M^{2}})}{{\bar R}^{2}(1)}] \cdot
 \nu^{2}(\frac{s}{M^{2}}) \int d^{2} b e^{- \Omega(s,b)}
    e^{- \frac{2 b^{2}}{R^{2}(\frac{s}{M^{2}})}}
\end{eqnarray}
As we have already seen  \cite{11} this integral can be evaluated
analytically yielding
\begin{eqnarray}
M^{2} \frac{d \sigma}{d M^2}\,\, =\,\,
 \frac{1}{8 \pi \alpha^{\prime}} \sigma_{PP}(M^{2})
 \nu^{2}(\frac{s}{M^{2}})
  ln [\frac{{\bar R}^{2}(\frac{s}{M^{2}})}{{\bar R}^{2}(1)}] \cdot
 a (\frac{1}{2\nu(s)})^{a} \gamma(a,2 \nu(s))
\end{eqnarray}
 where a is defined as
\beq
a(s,M^{2}) = 2\,\frac{ R^{2}(s)}{  R^{2}(\frac{s}{M^{2}})} \geq 2
\eeq
This allows us to conclude that
 \beq
< \vert D_{CD} \vert^{2} > =
 a (\frac{1}{2\nu(s)})^{a} \gamma(a,2 \nu(s))\,\,\underarrow{\alpha'
 ln s \gg R^{2}_{0}}
\,\,<\vert D_{SD} \vert^{2}>
\eeq
  Although the formal structure of Eq.(30)
appears to be identical to that of Eq. (9), there is a variance,
as $ a(s,M^{2}) $ is defined differently for SD and CD. For
SD, $ a \rightarrow $ 2 from below, whereas for CD $ a \rightarrow $ 2
from above. A general mapping of the damping factor
 as a function of a and $ \nu $
 is presented in Fig.4.
\par
    The asymptotic energy dependence of the integrated $ \sigma_{CD} $
is presented in Table I. Once again we note that for present day
energies  $ \nu \simeq $ 1.
  Calculated values of $<\vert D_{CD} \vert^{2}>$ are listed in
 Table II for some typical accelerator energies and are compared with the
damping factors calculated for other processes. We have  also
included a calculation
of $ < \vert D_{CD} (m^2_{\chi(3415)}) \vert^2 >$  for central
 $\chi_{c0}(3415)$ diffractive
production where we have assumed $R^2_0(\chi(3415)) \,=\, 1 GeV^{-2}$.

 Note that the above definition of $ a(s, M^2 )$, e.g Eq.(9),
corresponds to the case where the projectiles survive the collision intact.
For the case where projectiles are diffracted, $a ( s, M^2 )$
has a more complicated form, which does not converge as fast.

    To summarize our conclusions: the main result derived from
our calculations is that the energy scale at which the screening
corrections  for CD  saturate, are approximately the same
as that for SD \cite{11}. However, the damping factors for CD
 are  smaller than those for SD. As a
result the uncorrected cross sections calculated for various CD channels
 have to be scaled down
 by a factor of 3 - 3.5   at Tevatron energies and by a factor of  5
in the LHC energy range. This differs dramatically from the damping
calculated for the elastic amplitude.
\par
    Our second observation is that $ < \vert D_{i} \vert^{2} > $
 can be factored out from the $ M^{2} $ integration. The reason for
this is,  that $ a(s,M^{2}) $ is almost a constant due to its
 logarithimic dependence on $ M^{2} $. Consequently,
our results (damping factors)
 are applicable to the various cross sections for DPE
channels published recently [4-7]. These calculations were
  mainly made within the framework of the Low-Nussinov two gluon
 approximation of the Pomeron \cite{9,10}. Since, the amplitudes in this
 approximation appear to be well parametrized by an exponential
 in t-space, our method is applicable, and the amount of predicted
damping can easily be deduce from Fig.4, or Table II. To be more
specific we  present in Fig.5 a comparison between the published
 calculations of the DPE
contribution to heavy pair production \cite{6} and our the results
after damping corrections have been made.

\newpage
\vglue 0.5cm
%{ \elevenbf \noindent References}
\vglue 0.4cm

\newpage
\vglue 0.5cm
%{ \elevenbf \noindent FIGURE CAPTIONS}
\section*{Figure Captions}
\vglue 0.2cm
{}~~~
{\bf Fig.1:} The process of Double Pomeron Exchange (DPE).

{\bf Fig.2:} A typical lego plot for DPE.

{\bf Fig.3:} Relevant kinematics of our DPE calculation.

{\bf Fig.4:} A mapping of  damping factor as a function
of $ a(s,M^{2})$ and $ \nu(s)$.    \\

{\bf Fig.5:} DPE contribution to the cross section for heavy
quark pair production versus the CM energy of the collision.
Dashed lines denote the results of the uncorrected calculation
of Bialas and Szeremeta \cite{6}. Full lines are the results
after screening corrections have been made.

%{elevenbf \noindent TABLE CAPTIONS}
\section*{Table Captions}
\vglue 0.4cm
{}~~~~~{\bf Table I.} Behaviour of the asymptotic cross section for
uncorrected supercritical Pomeron model and the GLM model. \\

{\bf Table II.}  Damping factors at some typical accelarator energies \\
\newpage
\begin{center}

{\bf Table I.}

\begin{Large}

\begin{tabular}{|l|l|l|r}
   \hline
& Supercritical  & GLM  \\
&   Pomeron      &     \\   \hline
$\sigma_{tot}$  & $ s^{\Delta} $ & $ ln^{2} \frac{s}{s_{0}} $ \\ \hline
$ \sigma_{el}$  & $ \frac{s^{2 \Delta}}{ln \frac{s}{s_{0}}} $  &
$ ln ^{2} \frac{s}{s_{0}} $  \\  \hline
$ \sigma_{SD}$  & $ \frac{s^{2 \Delta}}{ ln \frac{s}{s_0}} $  &
$ ln \frac{s}{s_{0}} $  \\  \hline
$ \sigma_{CD}$  & $ \frac{s^{2 \Delta}}{ln \frac{s}{s_0}} $  &
$ ln \frac{s}{s_{0}} $  \\  \hline
$ \frac{\sigma_{el}}{\sigma_{tot}}$  &
 $ \frac{s^{ \Delta}}{ ln \frac{s}{s_{0}}} $  & $ \frac{1}{2}$ \\ \hline
$\frac{\sigma_{diff}}{\sigma_{tot}}$ & $\frac{s^{\Delta}}{ln \frac{s}{s_0}}$
& $\frac{ln\frac{s}{s_{0}}}{s^{\Delta}}$ \\ \hline
\end{tabular}
\end{Large}

\vspace{40 mm}
{\bf Table II.}

\begin{tabular}{|l|l|l|l|l|l|l|r}
   \hline
$ \sqrt{s}\; $ GeV  & $ \nu $ & $ < \vert D_{tot} \vert^{2} > $ &
$ < \vert D_{el} \vert ^{2} > $ & $ < \vert D_{SD} \vert ^{2} > $ &
$ < \vert D_{CD} \vert ^{2} > $ &
 $< \vert D_{CD}(m^2_{\chi(3415)}) \vert^2>$
 \\ \hline
180 & 0.76 & 0.838  & 0.626 & 0.446 & 0.344 & 0.356 \\ \hline
540 & 0.86 & 0.820  & 0.592 & 0.402 & 0.337 & 0.328 \\ \hline
1800 & 0.94 & 0.807  & 0.566 & 0.369 & 0.309 & 0.301 \\ \hline
\hline
 a = 2 & 1.00&0.797& 0.548&0.296&0.296 & 0.296\\ \hline
\hline
8000 & 1.17 & 0.770  & 0.501 & 0.295 & 0.243 & 0.232  \\ \hline
1400 & 1.20 & 0.765  & 0.494 & 0.289 & 0.234 & 0.223 \\ \hline
40000 & 1.31 & 0.749  & 0.467 & 0.262 & 0.208 & 0.201 \\ \hline
\end{tabular}
\end{center}
\end{document}